\documentclass[jkps,preprint,fleqn,showkeys]{revtex4}
\usepackage[pdftex]{graphicx}

\usepackage{amssymb}
\usepackage{amsmath}
\usepackage{bm}
\usepackage{color}
\usepackage{kotex} 

\usepackage[most]{tcolorbox}
\usepackage{dirtytalk}

\begin{document}
\setcounter{page}{0}
\title[]{Revealing role of the Korean Physics Society with keyword co-occurrence network}

\author{Seonbin \surname{Jo}}
\affiliation{Department of Physics, Pohang University of Science and Technology, 37673 Pohang, Korea}

\author{Chanung \surname{Park}}
\affiliation{Division of Social Data Science, Pohang University of Science and Technology, 37673 Pohang, Korea}

\author{Jungwoo \surname{Lee}}
\affiliation{Division of Social Data Science, Pohang University of Science and Technology, 37673 Pohang, Korea}

\author{Jisung \surname{Yoon}}
\affiliation{Department of Industrial and Management Engineering, Pohang University of Science and Technology, 37673 Pohang, Korea}

\author{Woo-Sung \surname{Jung}}
\email{wsjung@postech.ac.kr}
\affiliation{Department of Physics, Pohang University of Science and Technology, 37673 Pohang, Korea}
\affiliation{Division of Social Data Science, Pohang University of Science and Technology, 37673 Pohang, Korea}
\affiliation{Department of Industrial and Management Engineering, Pohang University of Science and Technology, 37673 Pohang, Korea}
\affiliation{Graduate School of Artificial Intelligence, Pohang University of Science and Technology, 37673 Pohang, Korea}


\date{Received \today}

\begin{abstract}
Science and society inevitably interact with each other and evolve together. Studying the trend of science helps recognize leading topics significant for research and establish better policies to allocate funds efficiently. Scholarly societies such as the Korean Physics Society (KPS) also play an important role in the history of science. Figuring out the role of these scholarly societies motivate our research related with our society since societies pay attention to improve our society.
Although several studies try to capture the trend of science leveraging scientific documents such as paper or patents, but these studies limited their research scope only to the academic world, neglecting the interaction with society. Here we try to understand the trend of science along with society using a public magazine named \emph{Physics and High Technology}, published by the Korean Physics Society (KPS). 
We build keyword co-occurrence networks for each time period and applied community detection to capture the keyword structure and tracked the structure's evolution. In the networks, a research-related cluster is consistently dominant over time, and sub-clusters of the research-related cluster divide into  various fields of physics, implying specialization of the physics discipline. Also, we found that education and policy clusters appear consistently, revealing the KPS's contribution to science and society. Furthermore, we applied PageRank algorithm to selected keywords (``semiconductor'', ``woman'', ``evading''...) to investigate the temporal change of the importance of keywords in the network. For example, the importance of the keyword ``woman'' increases as time goes by, indicating that academia also pays attention to gender issues reflecting the social movement in recent years.

\end{abstract}


\maketitle

\section{INTRODUCTION}

Modern science evolves in a complex way\cite{ref_1_science_evolve}. Entangled subfields of science evolve together due to an increase in interdisciplinary studies\cite{ref_2_rapid_intersection}. Figuring out this complicated evolution of science and underlying knowledge structure is an important task for researchers and policy makers. Researchers can search new research topics and create technology forcasts\cite{ref_3_forcast}. Policymakers can allocate limited resources efficiently\cite{ref_4_budget}. Furthermore, this knowledge structure strengthens our intuition about science.

Scholarly societies also played an important role in the history of science\cite{ref_5_importance_of_society}. Academic conferences hosted by such scholarly societies facilitate as a place for communication and debate between scientists. They also publish their own journals to improve certain scientific fields. For example, physical review letters are published by American Physical Society. The role of scholarly societies is not limited to research. They also focus on education, science policies, minorities, and the careers of scientists.

Numerous studies have succeeded to investigate trend of science\cite{ref_6_science_of_science, ref_7_nuclear_hong, ref_8_citation_network}. These studies concentrated on the structure and evolution of science without considering other factors such as social background. This is because previous studies pursue professional scientific articles, which have the purpose of delivering novel messages based on solid results\cite{ref_9_role_of_professional_articles}. Such scientific articles are not suitable to observe the trend of science from a different point of view. Furthermore, in order to investigate history of scholarly societies, we need to select dataset that includes societies opinions, their social projects, not limited in scientific era.

Here, our goal was to trace the history of the KPS(Korean Physical Society), a Korean scholarly society founded in 1952. For 60 years, the KPS has made efforts to disseminate physics in Korea and contributed to the advancement of science. To capture this effort, We select the public magazine \emph{Physics and High Technology} as our dataset to capture diverse features of the KPS. The \emph{Physics and High Technology} is published by the KPS. Traditionally, science magazines serve as a vigorous communication channel between scientists and the public. The public learns emerging topics in science and society. Authors show their expertise about science and express their opinions about current social issues. \emph{Physics and High Technology} has also these properties. The articles of \emph{Physics and High Technology} are written by professional authors who majored in physics. Articles cover qualified information about physics and modern technology, but they designate the public as the main reader. Therefore, the magazine's content is tailored to the public and their demand. For example, editorials include opinions of scientists about controversial topics in society, such as physical education policy in Korea. We expected that this data will deal with more diverse topics and issues than scientific articles.

Network is a useful tool to represent and investigate complex structures\cite{ref_12_network}. Keyword co-occurrence networks, in particular, is an appropriate method to reveal knowledge structure in a set of scientific corpora\cite{ref_13_keyword_co_occurrence_network}. For example, keyword co-occurrence networks are applied to study the knowledge structure of physics, technology, and patents\cite{ref_14_network_application_1_you, ref_15_network_application_2_yang, ref_16_network_application_3_patent}. Also, constructing a keyword co-occurrence network allows us to use various network analysis methods, such as community detection, centrality, PageRank, and other network properties.

We constructed keyword co-occurrence networks from articles of \emph{Physics and High Technology} to observe interactions between science and society. We also divided data at regular time intervals and built networks. This allowed us to see temporal changes in network properties. Community detection was implemented after constructing networks. Network clusters from community detection represented major topics dealt in \emph{Physics and High Technology}. PageRank analysis was also applied to track the importance of each keyword over time

As the brief result, we observed that keyword co-occurrence network has three major clusters. These three major clusters represent topics that the KPS discuss via \emph{Physics and High Technology}. There are research-related cluster, policy/education-related cluster, and KPS news related cluster. Each cluster has fine structures. We also saw that these fine structure changed depend on time. PageRank analysis showed us when single keyword becomes important in the keyword co-occurrence network. From this information, we glanced that the KPS has also intention to our society, not sorely focused on academic domain.

\section{METHOD}

\begin{figure}
\includegraphics[width=16cm]{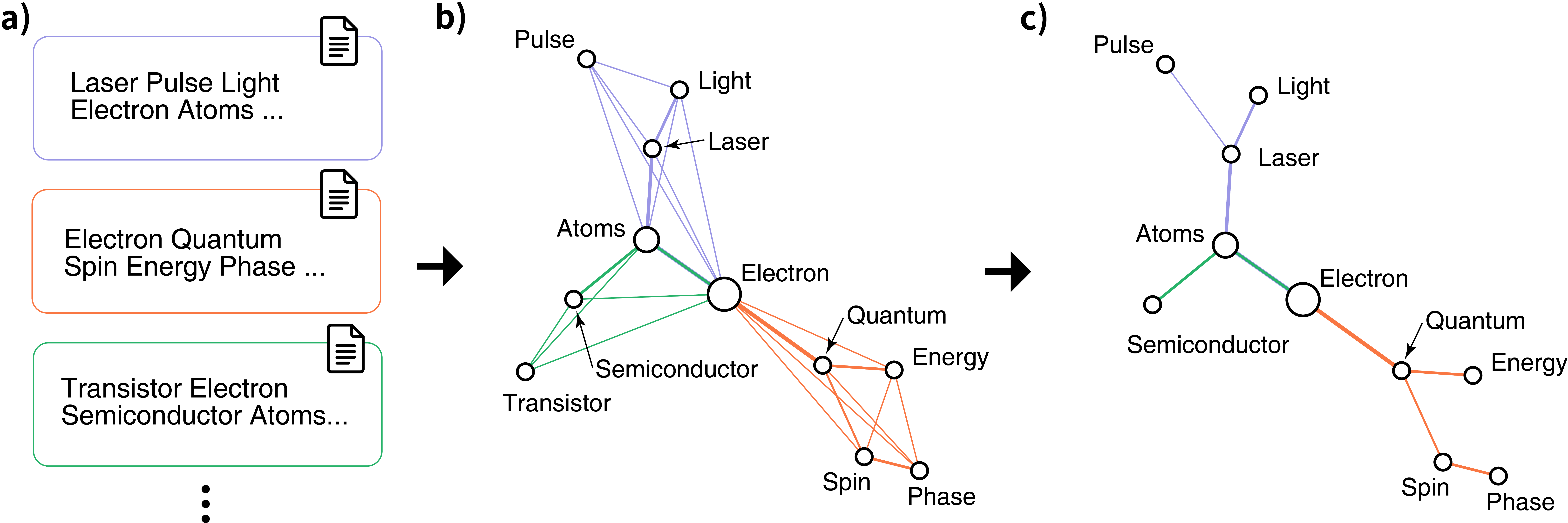}
\caption{Overall process to construct keyword co-occurrence network.
a) An example of preprocessed keywords from articles. Each colored block represents an article from \emph{Physics and High Technology}. Sentences are tokenized with Khaiii. TF-IDF threshold (0.05) is applied to each word. Arbitrary keywords are selected for visualization in a). 
b) An example of a keyword co-occurrence network. Every keyword in the corpus is a node. Nodes in the same article are fully connected (with undirected edges) and the edges are represented in an identical color. The size of a node represents the number of occurrences of the word, and the edge weight represents the number of co-occurrences between nodes. Edge weights are determined arbitrarily for visualization in b).
c) An example of a network after extracting the backbone. Keyword co-occurrence networks made by a bottom-up approach. A bottom-up approach uses all keywords from the corpus, making the network heavily dense. To extract useful information, network backbone extraction should be applied to remove trivial edges and isolated nodes.}
\label{fig1_overall_method}
\end{figure}

We selected 30 years of public magazine articles of \emph{Physics and High Technology} to trace the history of the KPS. This corpus contains 5,429 articles, 8,004,145 words, and 249,836 distinct words. Since the magazine's purpose is to introduce physics to the public, \emph{Physics and High Technology} contains not only information about physics itself, but also articles that deal with issues and events surrounding the KPS and concerns in society. These features of \emph{Physics and High Technology} allow us to analyze the flow of physics with the history of the KPS. We built a keyword co-occurrence network from each article in a bottom-up approach to see the structure of keywords. The overall procedure is presented in Figure \ref{fig1_overall_method}

A preprocessing procedure to clean the text of the corpus is needed before creating a keyword co-occurrence network. First of all, Khaiii (Kakao Hangul Analyzer III) was used to tokenize and morpheme tagging words. A local whitelist dictionary made by the KPS was also used to screen physical terminologies. Using the tokenizer, we saved separate words from sentences and part of speeches. From these morpheme tags, we selected words that are meaningful by themselves. For example, pronouns only have a contextual meaning and should be excluded. Nouns, verbs, and foreign languages were extracted as meaningful words from the corpus. Next, we selected important words from the remaining words. Words that appeared less than 5 times were removed. Also, we used TF-IDF (Term Frequency-Inverse Document Frequency), which is a relative variable that indicates the importance of words in an article, to gather essential words from the corpus. We excluded words below the TF-IDF threshold (0.05). We also tested different TF-IDF threshold values to see that this select might be distort the dataset. We used TF-IDF threshold 0.1, 0.15, and 0.2. As the result, we checked that main result is still held with different TF-IDF variables.

We constructed a keyword co-occurrence network with the preprocessed words. Every word is a node. If two distinct words appear in the same article, we added a weighted edge between the nodes. The size of the node represents the number of words that appear, and the weight of the edge represents the number of articles that contain both words. A network constructed following the method above is heavily dense because every word is a node, and nodes in an article form a complete subgraph, making it difficult to understand the keyword structure. A network backbone extract algorithm was applied to remove relatively irrelevant edges with a threshold $\alpha = 0.05$\cite{ref_19_backbone}. Isolated nodes were also removed immediately. We also created keyword co-occurrence networks with a regular time interval to see the temporal change of the network structure.

After constructing a keyword co-occurrence network, we used a community detection algorithm to figure out the underlying keyword structure behind \emph{Physics and High Technology} with a constant resolution parameter $\mu = 1$\cite{ref_20_louvain}. Each detected cluster was named manually by analyzing the meaning of keywords in the cluster. We also applied the algorithm to temporal networks to observe network structure change. The advantage of constructing keyword co-occurrence networks is that it allows us to apply general network analysis methods to analyze text data. A powerful method is PageRank\cite{ref_21_pagerank}. PageRank indicates the importance of a node in a network. We used PageRank to find important nodes and their time variance. We also compared with different centrality such as eigenvalue centrality and closeness. We observe that the tendency of PageRank and other centrality measures are matched each other.

\section{RESULT}

\begin{figure*}
\includegraphics[width=\textwidth]{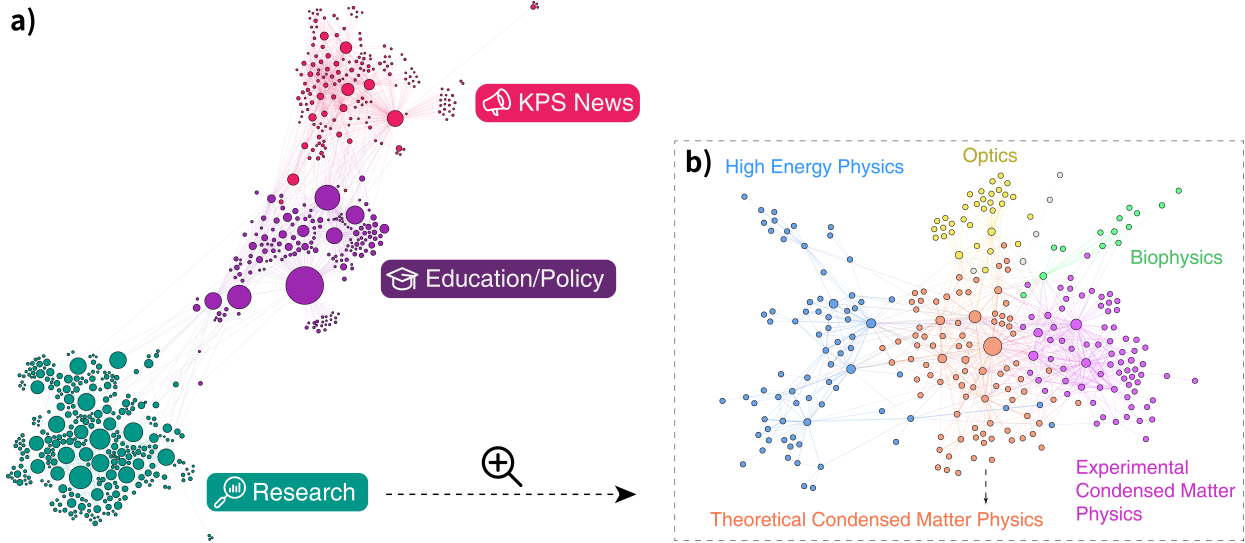}
\caption{Keyword co-occurrence network built with 30 years of \emph{Physics and High Technology} articles. Several minor clusters are omitted. a) Three major clusters appear in the network: Research-related cluster, education/policy-related cluster and KPS news-related cluster. Each cluster represents large topics that the KPS was willing to share with the public through \emph{Physics and High Technology}. b) Subclusters of research-related cluster. Each major cluster are also divided into subclusters. Subclusters represent fields of physics that the KPS and \emph{Physics and High Technology} handled.}
\label{fig2}
\end{figure*}

First, we investigated the keyword co-occurrence network constructed using all 30 years of articles, which is illustrated in Figure \ref{fig2}. We could observe three big clusters: research-related cluster, policy/education-related cluster, and KPS news-related cluster. These clusters represent the role of \emph{Physics and High Technology} as a communication channel of the KPS.

Research-related cluster includes keywords about physics. For example, electrons, energy, and plasma are in the research-related cluster. This cluster consists of words from science articles handled in \emph{Physics and High Technology} for 30 years. These articles were written by professional physicists affiliated to the KPS. The authors selected the topics of articles as their research topics to disseminate physics to the public. Therefore, the research-related cluster indicate the interest of the KPS for 30 years.

The policy/education-related cluster is an interesting cluster in the network. This cluster includes keywords such as research, government, policy, and education. The words in this cluster is from not only articles about policy/education, but also research articles. For example, ``semiconductors are important to society due to its industrial application'' Is a sentence in a research article but influences policy/education-related cluster. This cluster shows that the KPS is concerned about our society and not limited to the scholarly domain.

KPS news-related cluster includes keywords about events of the KPS, such as the result of regular meetings, the announcement of new associated members, and conference reports. We did not investigate further about this cluster because keywords appeared temporarily and did not form interesting structures.

\begin{figure}
\includegraphics[width=9.5cm]{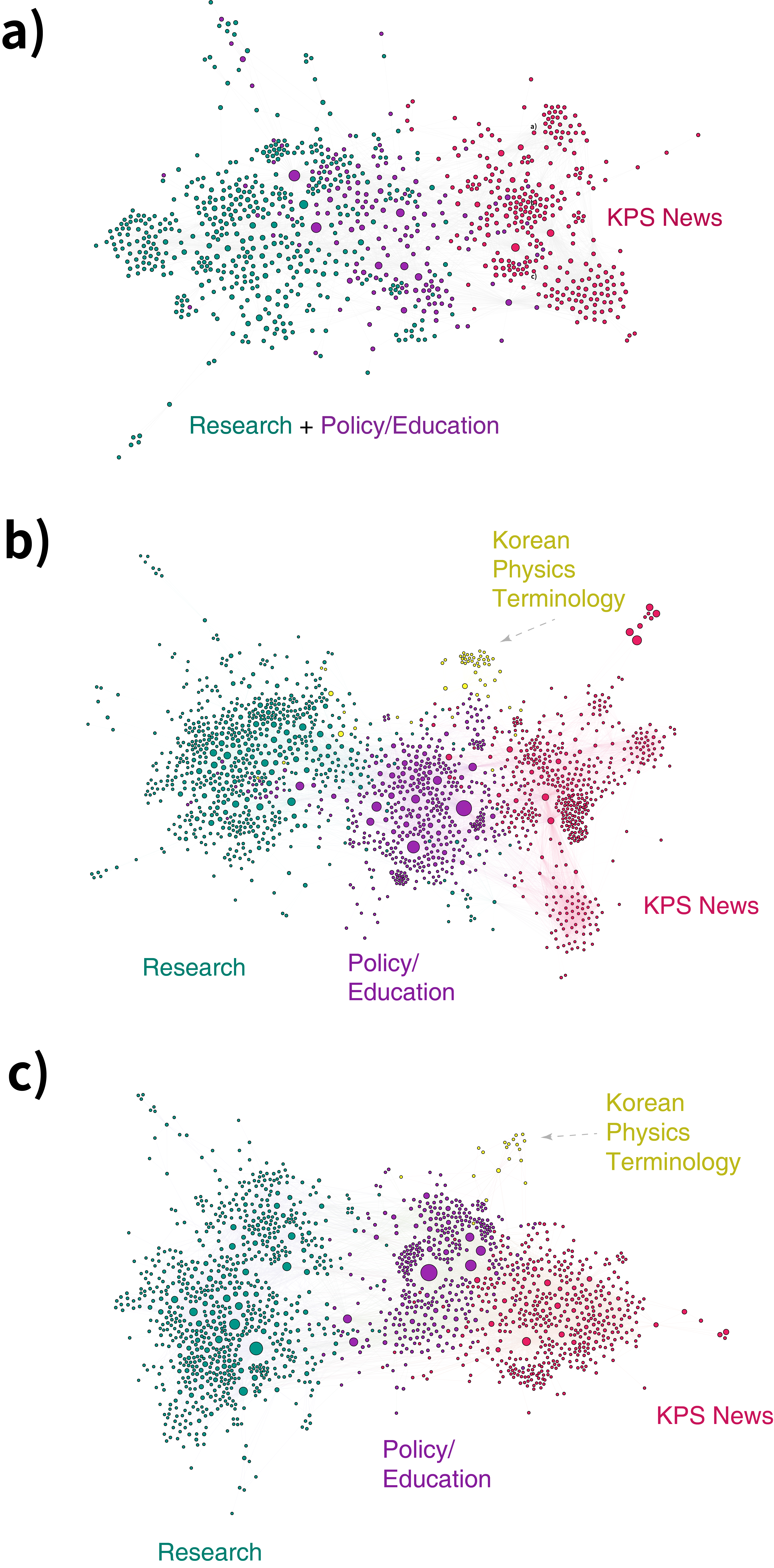}
\caption{Keyword network built with different time domain. a) 1992-1996, b) 1997-2001 c) 2002-2006. Education/policy-related cluster is within research-related cluster in figure a), but we could observe the separation of education/policy-related cluster from research-related cluster in figure b), indicating the emergence of a separate education/policy-related cluster. Mean cluster distance between networks in also increased from b) 3.11 to c) 3.39, supporting the distinction of clusters. The Korean government started to focus on science policy in the early 2000s, and we assumed this affected the content of the KPS.} 
\label{fig3}
\end{figure}

Each major cluster is also divided into subclusters. Research related cluster is divided into subdomain of physics, which is also illustrated on Figure \ref{fig3} b). Experimental condensed matter physics and its applications has the largest proportion, theoretical condensed matter physics is second largest, and high energy physics and astrophysics is third. Other subclusters such as optics, biophysics, and (quantum) information/communication are also in the research-related cluster. However, these subcluster proportions do not exactly correspond to the real proportions of research areas in the KPS. For instance, the proportion of astrophysics is significantly larger than real research domain. This is because some topics are selected based on the public's interest to popularize physics. Therefore, this proportion of subclusters may include public demand of physical fields or curiosity. Here we could see the effort of the KPS to try to solve the inquisitiveness of the public and the dissemination of physics itself.

Policy/education-related cluster is divided into two simple subclusters: the policy cluster and the education subcluster. Policy subcluster includes keyword related with science policy such as government funding policies and research topics related to economic or social problem. Education cluster includes keywords related to education, especially those related to physics in high school. For example, college admission policies have changed rapidly in Korea. This directly affects the physics curriculum in high school and the KPS shares opinions about these changes to educate the next generation of physicists. These efforts are captured in policy/education cluster.

The keyword co-occurrence network and its subclusters structure have changed over time because the trends of science and its surrounding environment, such as social events, also changed. We set 5-year time windows to divide our corpus and created keyword co-occurrence networks to capture the underlying temporal structure. Community detection method is also applied to each network.

Three major clusters that appear in the network above did not exist in the early era of the network. Initially, there are only research-related cluster and KPS news-related cluster. However, the policy/education-related cluster appears in the next time step, and we could observe separation between the policy/education-related cluster and research-related cluster. We measured the average minimum distance between nodes in each cluster, which is illustrated in Figure \ref{fig3}. The average distance between clusters changed from b) 3.11 to c) 3.39, indicating separation between clusters. This means that in the early era, authors mentioned research topics and science policy and education topics simultaneously in their articles However, the importance of science policy and education has arisen over time and the Korean government started to financially support this field from 1990s\cite{ref_21_korean_policy}. At the same time, the KPS started to handle science policy and education as a separate academic field.

We also observed one of the the KPS's achievements. As shown in Figure \ref{fig3}, there is a small subcluster around 2000s that could not be detected on all time networks due to coarse graining. This small cluster includes keywords about developing Korean physics terminology. In the 1980s, the Korean physics community used Chinese characters and English in both articles and papers. This confused the public's understanding of physics and disrupted communication between physicists. Therefore, the KPS has tried to unify this terminology in pure Korean, and this effort is observed in the subcluster.

\begin{figure}
\includegraphics[width=\textwidth]{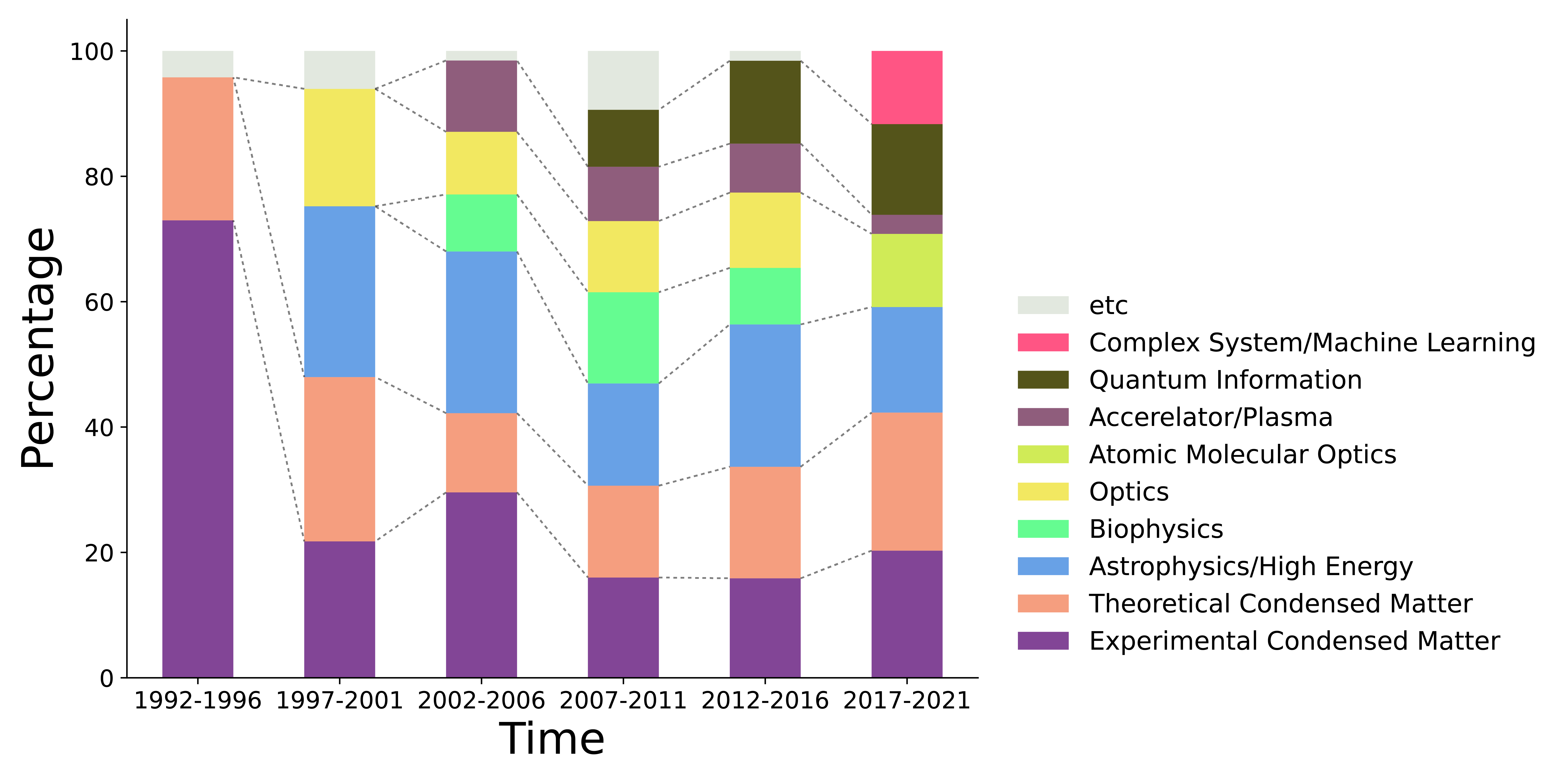}
\caption{Proportional changes of subclusters in research related cluster. Subfields that \emph{Physics and High Technology} handled have diverged over time.}
\label{fig4}
\end{figure}

Subclusters of research-related cluster have varied diversely, which represents the trend of fields of physics and their public interest. This is illustrated in Figure \ref{fig4} In the early era of the 1990s, condensed matter physics accounted for most of the research related cluster. This dominance ended in the late 1990s. Experimental and theoretical condensed matter physics is still a major subcluster, but we could observe optics and high energy physics/astrophysics clusters. Subclusters became more diversified after the early 2000s. Accelerator physics, plasma, and biophysics clusters appear in this era. There are several articles mentioned about the Hanaro accelerator at this time, which is the big science of Korea's government. Quantum information science also arises after the late 2000s, which is when the field of quantum information started to draw attention. The portion of quantum information science increases gradually, and this tendency continues until the late 2020s. In late 2020s, optics and biophysics cluster merges to a single field, atomic molecular and optical physics (AMO). Complex systems/machine learning clusters also emerge in the late 2020s.

\begin{figure*}
\includegraphics[width=\textwidth]{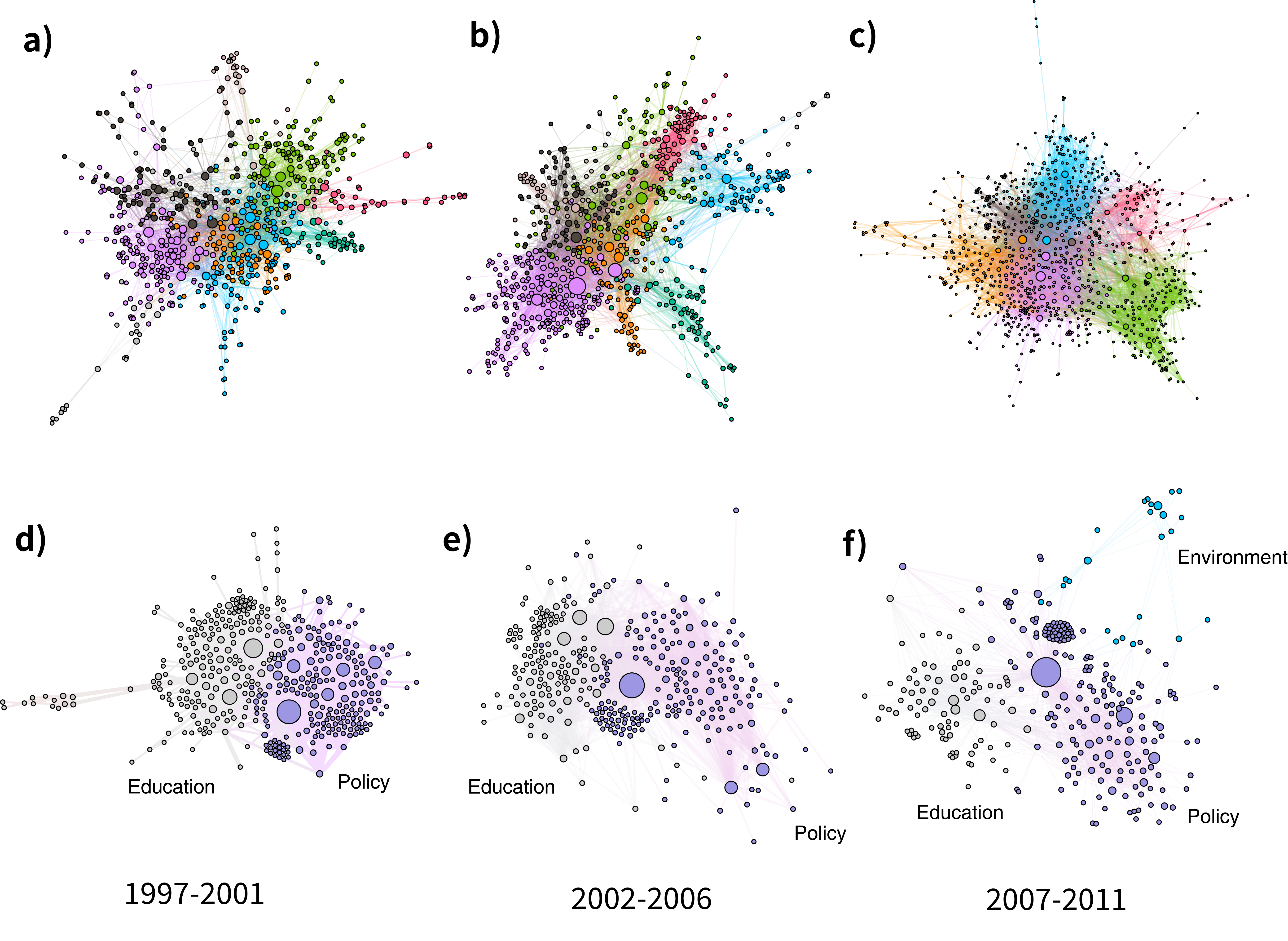}
\caption{Example of subclusters change in research-related clusters a), b), c) and policy/education-related cluster d), e), f). Cluster labels are omitted in figure a), b) and c) due to spatial problem. In research-related cluster, we could observe that each field changes diversely. However, in policy/education related cluster, subclusters are rather static than research related cluster.}
\label{fig5}
\end{figure*}

Subcluster of policy/education cluster has a simple structure compared to research related cluster. Policy/education cluster is divided into a policy cluster and an education cluster. This is illustrated in Figure \ref{fig5}. Although, environment cluster appears in Figure \ref{fig5} f), policy and education subcluster is dominant in all time era. Also, keywords of these subclusters do not change dramatically. This shows that the KPS and \emph{Physics and High Technology} have consistent opinions about policy and education over time.


\begin{figure*}
\includegraphics[width=\textwidth]{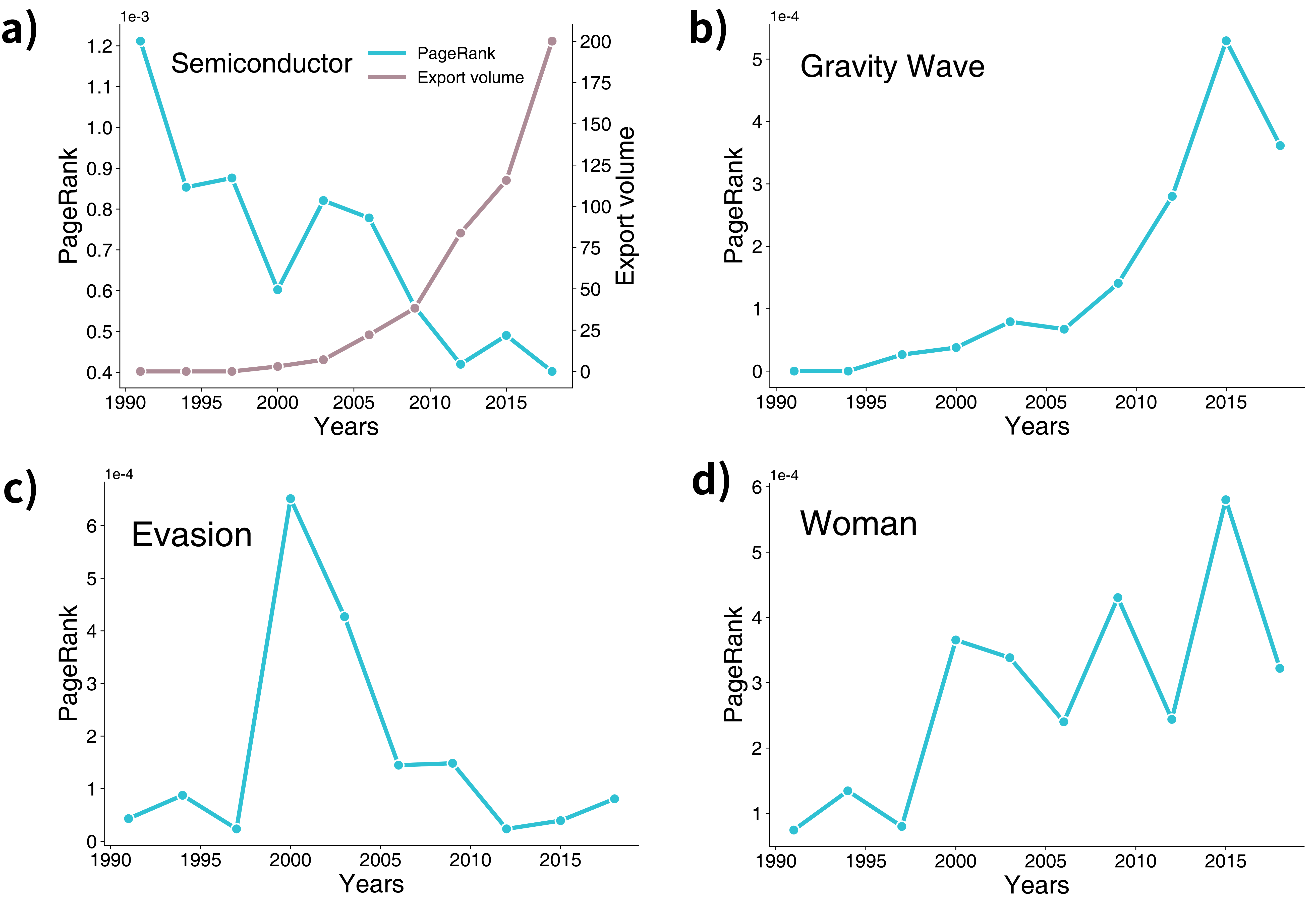}
\caption{PageRank changes of selected keywords. PageRank shows the importance of a node in the network. We could observe through keywords ``evasion'', and ``woman'' that the KPS has interest in our society as well.}
\label{fig6}
\end{figure*}

We also observed how the importance of a single keyword changes over time. We trace the change of PageRank of a single keyword, which represents the importance of a node in a network. We chose a 3-year time window to construct the keyword network to see changes more precisely. The network backbone extract method is not applied in this case due to the reduction of corpus. We selected ``semiconductor,'' ``gravity wave,'' evasion,'' and ``woman'' as keywords for our investigation The first two keywords are for research trends, and the others are for the social activity of the KPS.

We could observe interesting results from the keyword ``semiconductor''. This is shown in Figure \ref{fig6} a). We chose this keyword because condensed matter physics has always been the dominant subcluster of the research-related cluster, and semiconductors are a major application of condensed matter physics. However, PageRank of ``semiconductor'' gradually decreases. We compared this with the export volume of semiconductors from Korea. The export volume of semiconductors increased over time unlike the PageRank of ``semiconductor''. We assume the export volume of semiconductors represent the importance of industrial application of semiconductors in Korea, whereas \emph{Physics and High Technology} focuses on pure physics and the leading edge of technology. Therefore, we suggested the dimension of importance of semiconductors shifted from a pure science and technology level to an industrial level. This phenomenon is also explained by the linear model of innovation.

The Nobel Prize is always an issue to both scientists and the general public. Several articles in \emph{Physics and High Technology} focus on the Nobel prize and explain evolved physics. Nobel Prize-related keywords appeared frequently in articles about policy and education, indicating that the public is interested in them as well. However, these keywords did not form distinct subclusters. We used several keywords that lead to the Nobel prize, such as ``topological insulator,''``graphene,'' ``gravity wave'' and  ``black hole ''. We observed that PageRank reached its peak when a keyword lead to the nomination of a Nobel prize. Also, PageRank reached its peak gradually. This means that the KPS has been focusing on important topics in physics which could lead to the Nobel prize.

The KPS has been concerned with social problems related to science as well. This is implied in the keywords ``evasion'' and ``woman''. The PageRank of ``evasion'' suddenly reached its peak in the early 2000s. This is illustrated on Figure \ref{fig6} c) and d) At that time, the STEM evasion problem was a controversial issue in Korea\cite{ref_22_evasion}. Students were disappointed by the scientific fraud and embezzlement of a renowned scholar, and the number of students entering graduate school significantly dropped at that time. The KPS also handled this problem seriously and expressed opinions through \emph{Physics and High Technology}. The KPS also focused on gender issues, which could be observed through the keyword ``woman''. Gender inequality of STEM is a major problem in Korean society. We chose this keyword to see if this problem also occurs in articles. As a result, the PageRank of ``woman'' gradually increased over time, indicating that the KPS was also concerned about the gender issue. For example, the KPS has operated a women's committee and other events for female physicists and students, such as a physics camp for female high school students to encourage them to study physics. These KPS's efforts to improve our society is not ended without success. For instance of ``evasion'', There are articles that KPS's opinion is reflected to high school physics curriculum to encourage student to study physics. Furthermore, in case of Korean Physics terminology, We could find article that usage of Korean physics terminology is increased in \textit{New Physics: Sae Mulli}. The success of KPS's effort is also contained in \textit{Physics and High Technology}

\section{SUMMARY}

We investigated the interest and role of the KPS through a keyword co-occurrence network with 30 years of public magazine articles in \emph{Physics and High Technology}. The Keyword co-occurrence network has three clusters: research-related cluster, policy/education-related cluster, and the KPS news-related cluster. These clusters represent the role of \emph{Physics and High Technology} that the KPS wants to deliver to the public. Each major cluster also has subclusters. Subclusters research-related clusters, for example, are fields or physics that the KPS handled. The structures of these cluster structures change over time. We observed a split between the policy/education-related cluster and the research -related cluster. Also, we could find a small cluster that indicates the contribution of the KPS in unifying Korean physics terminology. Research related clusters change variously, which means that physics topics from \emph{Physics and High Technology} becomes diverse. Policy/education-related cluster are rather static than research related clusters. Finally, we traced the importance of a single keyword via PageRank and identified interactions between society and the KPS.

We showed that the KPS traces through public articles rather than professional physics research papers. Keyword co-occurrence network of these non-professional datasets allowed us to observe opinions and contributions of the KPS to society. We analyzed the emergence of policy/education-related clusters, traces of PageRank of non-scientific keywords, clusters of Korean physics terminology, and many more observations. Although some proportion of research subclusters do not correspond to actual research areas in physics, we assumed this difference indicates the public's interest.

The limitation of this approach is that we could not figure out any contextual results from articles. For example, we only know that \emph{Physics and High Technology} discuss education. However, we could not distinguish whether the opinion of the KPS about education is positive or negative because every contextual meaning is omitted. Extracting contextual meaning from the corpus may provide us more accurate results, such as finding the differences in static policy/education subclusters. Also, the responses of the public are not included in this research. If we compare our research with other media that include the public’s opinions and responses about the KPS or other social events for future research, we may be able check the validity of our work.

\section{ACKNOWLEDGEMENT}
This research was motivated by request of analysis of \emph{Physics and High Technology} from KPS. Also, This research was supported by the National Research Foundation (project number : 2021R1F1A106303012). This work was supported by Institute of Information \& communications Technology Planning \& Evaluation (IITP) grant funded by the Korea government(MSIT) (No.2019-0-01906, Artificial Intelligence Graduate School Program(POSTECH))


\end{document}